\documentclass[sigconf,balance=false]{acmart}
\usepackage{float}
\usepackage{flafter}
\usepackage{tabularx}
\usepackage{colortbl}
\usepackage{placeins}
\usepackage[skins,breakable]{tcolorbox}
\usepackage{algorithm}
\usepackage{algpseudocode}
\algrenewcommand{\algorithmicrequire}{\textbf{Input}}
\algrenewcommand{\algorithmicensure}{\textbf{Output}}

\graphicspath{{figures/}}
\setcopyright{none}
\acmConference[WWW '27]{The ACM Web Conference 2027}{May 10--14, 2027}{Dublin, Ireland}
\acmYear{2027}
\copyrightyear{2027}
\acmDOI{}
\acmISBN{}
\title[PCG-RAG for Complex Web Question Answering]{PCG-RAG: Progressive Compositional Grounding for Retrieval-Augmented Generation in Complex Web Question Answering}
\newcommand{\pcg}{PCG-RAG}

\newcommand{\Pa}{\operatorname{Pa}}

\definecolor{TableInk}{HTML}{29485B}
\definecolor{TableHead}{HTML}{EAF1F6}
\definecolor{TableGroup}{HTML}{F3F6F8}
\definecolor{TableEmphasis}{HTML}{EAF4F1}
\newcommand{\tablestyle}{\small\setlength{\tabcolsep}{4pt}\renewcommand{\arraystretch}{1.08}\arrayrulecolor{TableInk}}
\newcommand{\tablehead}[1]{\textcolor{TableInk}{\textbf{#1}}}
\newcommand{\tokenpair}[2]{\makebox[3em][r]{#1}\,/\,\makebox[2.5em][r]{#2}}
\newcolumntype{L}{>{\raggedright\arraybackslash}X}
\newcolumntype{R}{>{\raggedleft\arraybackslash}X}
\definecolor{PromptInk}{HTML}{29485B}
\definecolor{PromptFrame}{HTML}{B8CAD6}
\definecolor{PromptBody}{HTML}{F5F8FA}
\definecolor{PromptMuted}{HTML}{526874}

\newtcolorbox{promptcard}[1]{%
  enhanced,
  breakable,
  lines before break=3,
  colback=PromptBody,
  colframe=PromptFrame,
  boxrule=0.45pt,
  arc=1.1mm,
  left=1.6mm,
  right=1.6mm,
  top=1.0mm,
  bottom=1.0mm,
  before skip=1.6mm,
  after skip=1.6mm,
  title={#1},
  colbacktitle=PromptInk,
  coltitle=white,
  fonttitle=\small\bfseries,
  toptitle=0.7mm,
  bottomtitle=0.7mm,
  titlerule=0pt,
  fontupper=\small,
}

\newcommand{\promptfield}[1]{%
  \par\noindent{\color{PromptInk}\bfseries #1}\enspace\ignorespaces
}
\newcommand{\promptvar}[1]{\texttt{#1}}
\newcommand{\appparagraph}[1]{\par\smallskip\noindent\textbf{#1}\quad}

\author{Linyao Zheng}
\affiliation{%
  \institution{Xi'an Jiaotong University}
  \city{Xi'an}
  \country{China}
}

\author{Xuhang Shi}
\affiliation{%
  \institution{Beijing University of Posts and Telecommunications}
  \city{Beijing}
  \country{China}
}

\author{Zhifang Mao}
\affiliation{%
  \institution{Beijing University of Posts and Telecommunications}
  \city{Beijing}
  \country{China}
}

\author{Chongqi Qu}
\affiliation{%
  \institution{Xi'an Jiaotong University}
  \city{Xi'an}
  \country{China}
}

\author{Xiuquan Hou}
\affiliation{%
  \institution{Xi'an Jiaotong University}
  \city{Xi'an}
  \country{China}
}
\renewcommand{\shortauthors}{Zheng et al.}

\begin{document}
\begin{abstract}
Complex Web question answering requires retrieval-augmented generation (RAG) systems to identify target entities from multi-clue descriptions and connect evidence scattered across long Web pages. Individual pages often support only some of the required semantic constraints, while clues discovered on one page guide subsequent searches on others. However, premature reliance on partially supported entity hypotheses can divert retrieval from the original information need. RAG systems must therefore use partial evidence to guide further searches while seeking the evidence needed to establish the intended entities. We propose PCG-RAG, a novel RAG framework based on Progressive Compositional Grounding (PCG). Progressive Evidence Acquisition directs subsequent searches toward missing evidence for candidate attributes and relations. Compositional Reasoning separates candidate-guided retrieval from evidence-based answering along a directed acyclic graph (DAG) of subquestions. Provisional names make dependent searches specific, while answering tasks use accepted results and unresolved identifying conditions, assessed against the original question. New passages connect the identifying clues to the requested relations, supporting evidence-based downstream answers. Evidence Grounding combines identifying and relational source passages with supported intermediate answers, connecting evidence across Web pages to the original conditions during final generation. On BrowseComp-Plus and WebWalkerQA, PCG-RAG improves accuracy over the strongest compared baselines by 5.43 and 5.15 percentage points, respectively.

\end{abstract}


\maketitle
\section{Introduction}
\label{sec:intro}

Complex Web question answering requires identifying entities from multiple clues and combining facts distributed across Web pages. Retrieval-augmented generation (RAG) acquires external evidence to answer these questions~\cite{lewis2020retrieval,fan2024survey}. A page matching part of an entity description provides a name that makes subsequent searches more specific, even when the remaining identifying conditions are unresolved. Entity identification therefore guides evidence discovery while depending on the evidence retrieved. We study this problem within the Search, Recommendation, and Retrieval-Augmented AI track.

Multi-step RAG uses intermediate findings to determine subsequent information needs. IRCoT interleaves retrieval with reasoning~\cite{trivedi2023ircot}, and ReAct incorporates observations into later actions~\cite{react}. Web search systems extend these processes through browsing and document reading~\cite{webgpt,webthinker,searcho1}. ReWOO, LQT, and LogicRAG organize dependencies so that intermediate results guide subsequent retrieval~\cite{xu2023rewoo,lqt,chen2026logicrag}. In entity-dependent Web QA, a name can be useful for search before all identifying conditions are established. Treating this search hypothesis as an established identity makes later relation answers depend on an unverified premise. Subsequent reasoning therefore needs the original conditions and identifying sources to assess whether a newly retrieved relation concerns the intended entity.

\begin{figure*}[t]
  \centering
  \includegraphics[width=0.94\textwidth]{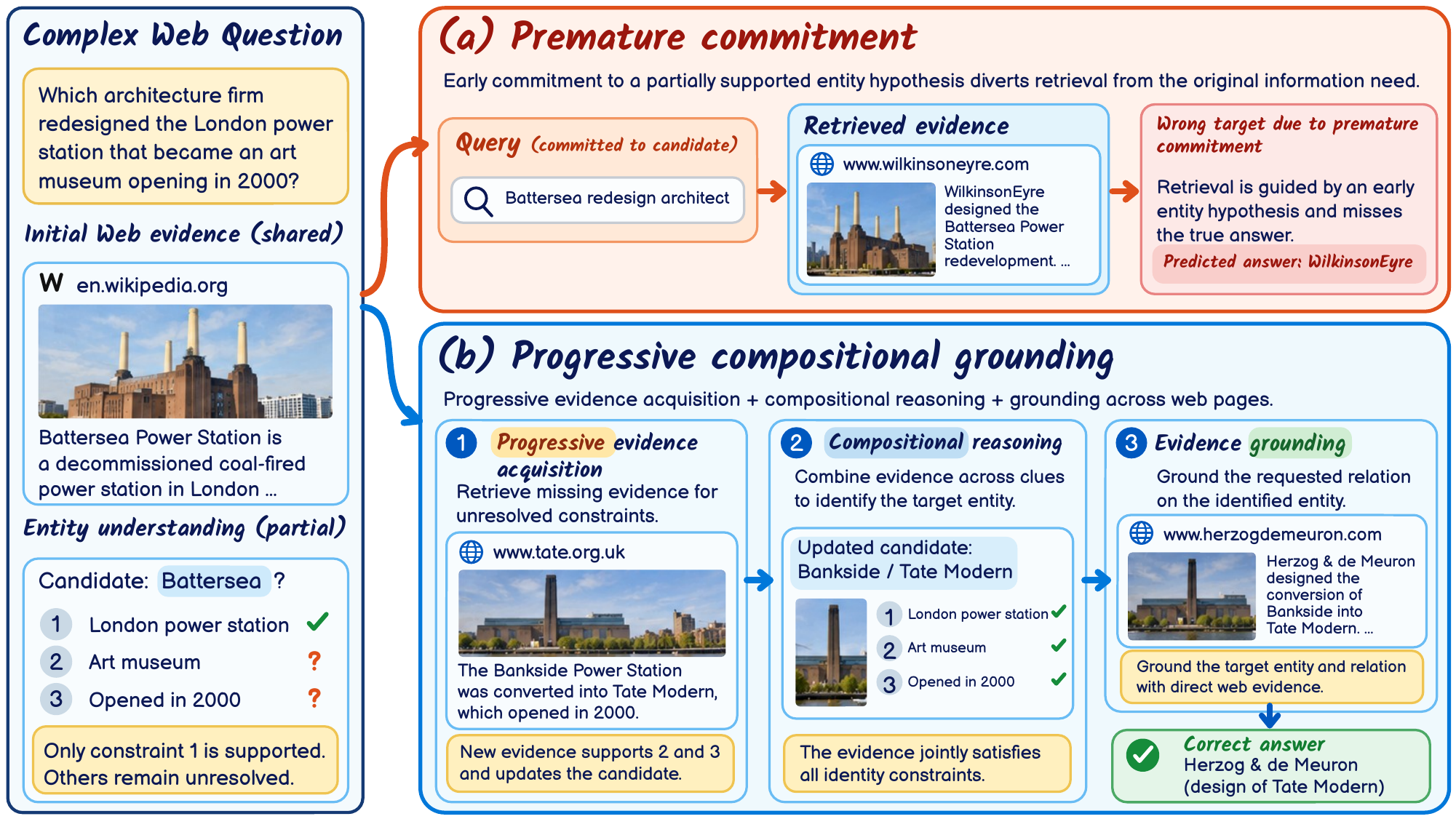}
  \caption{Two retrieval paths from a partial entity match. In (a), committing to Battersea directs the architect query toward a fact about the wrong subject. In (b), acquiring the missing identifying evidence establishes Bankside, whose conversion is attributed to Herzog \& de Meuron.}
  \Description{Both paths start with evidence that Battersea is a London power station, leaving the museum and opening year unresolved. The upper path searches for Battersea's redesign architect and returns WilkinsonEyre. The lower path retrieves evidence identifying Bankside as Tate Modern, opened in 2000, and combines it with a source attributing the conversion to Herzog and de Meuron.}
  \label{fig:problem}
\end{figure*}

Figure~\ref{fig:problem} illustrates how a partial identity directs the evidence obtained for an answer. Battersea matches the London power-station clue, leaving the museum and opening year unresolved. In panel (a), an architect query about Battersea returns WilkinsonEyre. This passage supports a fact about Battersea's redevelopment, while the question describes the site of an art museum opened in 2000. In panel (b), retrieval follows the unresolved conditions. Evidence from Tate identifies Bankside as the site of Tate Modern, which opened in 2000, and an architectural source attributes its conversion to Herzog \& de Meuron. Combining these sources connects the architectural fact to the entity specified by the question.

An entity judgment consequently shapes subsequent evidence discovery, while later passages can change the subject of dependent reasoning. A compositional answer must connect the resulting relation to the identifying conditions through evidence from these pages. \textbf{\emph{How can RAG couple entity identification with compositional reasoning as Web evidence is progressively acquired?}}

We propose \textbf{\pcg{}}, a RAG framework for Progressive Compositional Grounding that couples entity identification with compositional reasoning as Web evidence is acquired. \emph{Progressive Evidence Acquisition} uses partial entity matches as search hypotheses to retrieve missing identifying evidence. New passages refine these hypotheses and establish the entities needed for subsequent reasoning. \emph{Compositional Reasoning} uses candidate identifications to retrieve dependent relations and establishes each answer from sources connecting the relation to the original identifying conditions. \emph{Evidence Grounding} organizes the identifying and relational sources with their intermediate answers, connecting the evidence across Web pages to the original question during final generation.

The main contributions are as follows.
\begin{itemize}
\item We propose \emph{Progressive Evidence Acquisition}, which uses partial entity matches as search hypotheses to retrieve evidence for unresolved identifying conditions. By distinguishing these hypotheses from accepted answers, it allows newly acquired passages to refine entity identification and guide subsequent retrieval.

\item We propose an evidence-grounded approach to compositional reasoning over evolving entity identifications. \emph{Compositional Reasoning} uses candidate entities to make dependent searches specific while grounding their answers in the original identifying conditions. \emph{Evidence Grounding} combines the identifying and relational sources, connecting intermediate conclusions to the original question's conditions during final answer generation.

\item Experiments on BrowseComp-Plus and WebWalkerQA~\cite{browsecompplus,webwalker} show that \pcg{} improves accuracy by 5.43 and 5.15 percentage points over the strongest compared baselines, respectively.
\end{itemize}

\section{Related Work}
\label{sec:related}

\subsection{Web Search Agents}

Web search agents combine language-model reasoning with search and browsing. WebGPT studies browser-assisted question answering with human feedback~\cite{webgpt}, and ReAct interleaves reasoning and actions using environmental observations~\cite{react}. Mind2Web and WebArena establish evaluation settings for interaction across websites~\cite{mind2web,webarena}, while WebVoyager integrates multimodal page understanding into browsing~\cite{webvoyager}. These works develop the action selection and page interpretation capabilities needed to obtain Web information.

Deep-search methods further integrate evidence acquisition into extended reasoning. Search-o1 uses Reason-in-Documents to extract useful information before continuing reasoning~\cite{searcho1}, while WebThinker embeds search and navigation in a deep exploration tool~\cite{webthinker}. Q2Q reformulates queries to improve retrieval and ranking~\cite{q2q}, and DeepHone studies reranking choices and their token cost along search trajectories~\cite{deephone}. Complex Web QA also requires the search subject itself to be established from evidence. \pcg{} uses partially identified entities to guide retrieval of the remaining identifying conditions, then connects the resulting identity judgments to dependent relation queries.

\subsection{RAG for Web Question Answering}

Multi-step RAG turns intermediate information needs into queries over sparse and dense retrieval~\cite{robertson2009bm25,karpukhin2020dpr}. Self-Ask generates follow-up questions~\cite{press2023selfask}, and IRCoT interleaves retrieval with intermediate reasoning~\cite{trivedi2023ircot}. FLARE uses uncertain predicted text to trigger retrieval~\cite{flare}, while Self-RAG learns reflection tokens for retrieval and response assessment~\cite{selfrag}. MetaRAG and RetroRAG use reasoning monitoring and evidence reassessment to guide further acquisition~\cite{zhou2024metarag,xiao2025retrorag}. Dependencies can also be represented explicitly. Corpus graphs connect evidence across a collection~\cite{edge2024graphrag,jimenez2024hipporag}, while ReWOO binds observations to variables in a question-specific plan~\cite{xu2023rewoo}. LQT organizes atomic queries into a logical query tree through cost-guided dynamic programming~\cite{lqt}. LogicRAG executes an adaptive query dependency graph with rolling memory~\cite{chen2026logicrag}. Building on the connection between retrieval and intermediate reasoning, \pcg{} distinguishes provisional search subjects from accepted answers. Candidate entities guide dependent retrieval, while the original identifying conditions and acquired sources determine the support for each relation answer.

The resulting answers also depend on how their sources are organized. RADAR models document support and conflict to select evidence under retrieval corruption~\cite{radar}, while ALCE evaluates answer attribution through cited passages~\cite{alce}. Long-context studies show that passage position affects evidence use~\cite{lostmiddle}, and set-based selection considers the contribution of passage collections~\cite{lee2025setselection}. For entity-dependent Web questions, a relation passage must be connected to the sources identifying its subject. \pcg{} accumulates direct citations with the recorded sources of accepted parent answers and selects their unions under the generation budget. The generation context admits an intermediate answer only alongside its complete recorded citations, preserving the identity-to-relation source chain for final reasoning.

\section{Method}
\label{sec:method}

\begin{figure*}[t]
  \centering
  \includegraphics[width=0.90\textwidth]{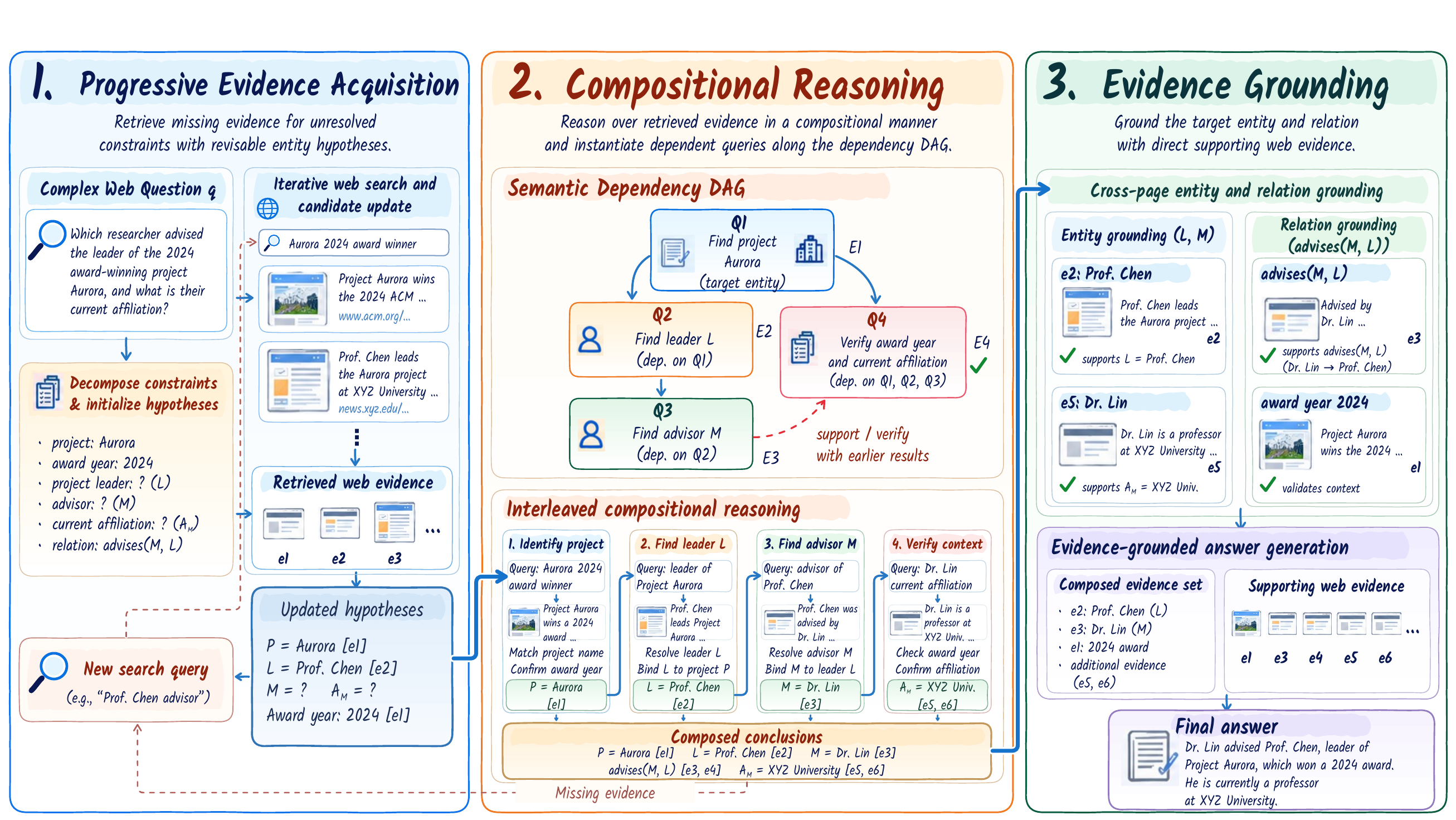}
  \caption{Overview of \pcg{}, illustrated with a question about a project leader's advisor and the advisor's current affiliation. Progressive evidence acquisition updates entity hypotheses as missing evidence is retrieved. Compositional reasoning instantiates dependent queries along the subquestion DAG. Identifying and relational passages jointly support final answer generation.}
  \Description{The left panel decomposes the question into project, award year, leader, advisor, and affiliation constraints and updates hypotheses through Web search. The middle panel follows a dependency DAG to identify Project Aurora, its leader Prof. Chen, the advisor Dr. Lin, and the advisor's current affiliation. The right panel combines passages supporting these identities and relations to generate an answer.}
  \label{fig:method}
\end{figure*}

\pcg{} couples entity identification with compositional reasoning through progressive Web retrieval. Evolving identifications guide dependent queries, while recorded sources connect identifying evidence to relation answers in final generation (Figure~\ref{fig:method}).

\subsection{Progressive Evidence Acquisition}
\label{sec:acquisition}

Progressive Evidence Acquisition uses partial entity matches to seek the evidence needed for identification. Given a Web question $Q$, the language model produces subquestion templates $q_i$, each with an output variable $x_i$. A candidate supplies concrete search terms, while the remaining identifying conditions specify what to investigate. In Figure~\ref{fig:problem}, the initial Battersea match leaves the museum and opening year unresolved. These conditions guide further retrieval beyond the power-station description.

Let $t$ index updates to retrieved evidence and subquestion results. The archive $W^t$ contains passage identifiers linked to their source spans, and $W_i^t\subseteq W^t$ denotes the passages visible for subquestion $i$. The model assesses these passages against $Q$ and the current task, proposing an answer $b_i^t$ with direct citations $J_i^t\subseteq W_i^t$ and a description of missing evidence $m_i^t$. Grounding links each judgment to the source passages cited by the model. Semantic support is assessed against the question's identifying conditions and target relation.

The assessment status $s_i^t$ expresses this evidence judgment. The model assigns \emph{supported} when it judges the answer established under the original conditions. A \emph{provisional} answer includes a cited candidate with a stated gap. \emph{Insufficient} indicates that the current passages establish no candidate. For an unresolved subquestion, the current candidate $h_i^t$ is updated after a valid assessment by
\begin{equation}
  h_i^{t+1}=\begin{cases}
    (b_i^t,J_i^t,m_i^t), & s_i^t=\texttt{provisional},\\
    \bot, & \text{otherwise}
  \end{cases}.
  \label{eq:candidate}
\end{equation}
Here $\bot$ denotes absence, and $c_i^t$ denotes the answer component of a nonempty candidate state $h_i^t$. The model combines this candidate with the missing conditions to formulate the next query, using available pages for focused reading and discovered links for navigation. Each acquisition expands the passage archive. New passages are considered with cited parent sources, allowing the next assessment to replace the candidate. Candidate updates direct further retrieval toward the remaining identifying conditions.

Acceptance requires a nonempty supported answer with aligned citations and no stated gap. The accepted answer is stored as $a_i^{t+1}$ with direct citations $S_i^{t+1}$. Algorithm~\ref{alg:acquisition} summarizes this acquisition process. Acceptance ends the attempt. Further acquisition requires remaining allowance and an admissible action, as detailed in Appendix~\ref{app:context}.

\begin{algorithm}[H]
  \caption{Progressive Evidence Acquisition}
  \label{alg:acquisition}
  \begin{algorithmic}[1]
    \Require $Q$, subquestion $i$, parent results, archive $W$, acquisition allowance
    \Ensure Answer with direct citations, candidate, status, updated $W$
    \State Retrieve initial passages using the current parent bindings
    \State Add retrieved passages to $W$
    \State Compose visible passages with upstream sources
    \Loop
      \State Assess visible evidence against $Q$ and the task
      \State Update the candidate and missing conditions
      \If{the answer meets the acceptance conditions}
        \State \Return accepted answer with its direct citations
      \EndIf
      \If{an acquisition stopping condition applies}
        \State \Return current candidate and evidence status
      \EndIf
      \State Acquire evidence for the remaining conditions
      \State Archive obtained passages and refresh visible context
    \EndLoop
  \end{algorithmic}
\end{algorithm}

\subsection{Compositional Reasoning}
\label{sec:composition}

An entity judgment determines the subject of subsequent relation retrieval. Compositional Reasoning uses these dependencies to guide searches and establishes relation answers from evidence matching the original identifying conditions. For $n$ subquestions, the fixed dependency DAG is
\begin{equation}
  \begin{aligned}
    G&=(V,E),\\
    V&=\{1,\ldots,n\},\\
    E&=\{(j,i)\in V\times V\mid x_j\text{ is an input to }q_i\}.
  \end{aligned}
  \label{eq:dag}
\end{equation}
Subquestions are indexed in topological order, so each edge satisfies $j<i$. The parent set is $\Pa(i)=\{j\in V\mid(j,i)\in E\}$. Each subquestion is attempted after its parents, including parents whose identities remain unresolved.

Let $R^t$ denote the subquestions with accepted answers $a_j^t$. For an unresolved parent, $d_j^t$ describes its identifying task with accepted ancestor answers substituted. The answering task uses accepted answers while preserving these unresolved conditions
\begin{equation}
  \begin{aligned}
    q_i^t&=q_i[x_j\leftarrow f_j^t]_{j\in\Pa(i)},\\
    f_j^t&=\begin{cases}
      a_j^t, & j\in R^t,\\
      d_j^t, & j\notin R^t
    \end{cases}.
  \end{aligned}
  \label{eq:bindings}
\end{equation}
For retrieval, set $\widetilde f_j^t=c_j^t$ when unresolved parent $j$ has a cited candidate, and $\widetilde f_j^t=f_j^t$ otherwise. These bindings define a retrieval seed
\begin{equation}
  \widetilde q_i^t=q_i[x_j\leftarrow\widetilde f_j^t]_{j\in\Pa(i)}.
  \label{eq:hypothesis}
\end{equation}
Candidate names make searches specific, while Equation~\ref{eq:bindings} preserves unresolved conditions for answering. For the advisor subquestion in Figure~\ref{fig:method}, a provisional leader name narrows search, while assessment still requires evidence connecting that leader to the specified project. The original question $Q$ remains part of every assessment, including when a parent answer has been accepted. A child can be accepted before its parents are resolved when its own sources establish both the required identity and relation. Appendix~\ref{app:prompts} gives the dependent-reasoning instructions.

The two task formulations make a candidate useful for search while keeping the answer grounded in the original identifying conditions. A newly retrieved page can provide both an identifying clue and the dependent relation. Parent citations supply evidence for conditions established on earlier pages. Each dependent answer thus connects the identifying and relational passages available for its task. Algorithm~\ref{alg:composition} follows the dependency order, with acquisition bounds specified in Appendix~\ref{app:context}.

\begin{algorithm}[H]
  \caption{Compositional Reasoning}
  \label{alg:composition}
  \begin{algorithmic}[1]
    \Require $Q$, acquisition allowances
    \Ensure Subquestion results and passage archive $W$
    \State Construct $G$ and retrieve initial passages for $Q$ into $W$
    \State Initialize accepted subquestions $R=\varnothing$
    \For{each subquestion $i$ in topological order}
      \State Form $q_i^t$ and $\widetilde q_i^t$ using Equations~\ref{eq:bindings}--\ref{eq:hypothesis}
      \State Solve $i$ with Algorithm~\ref{alg:acquisition} using $Q$, parent results, and $W$
      \State Record its answer, candidate, and cited sources
      \If{the answer is supported}
        \State Add $i$ to $R$
      \EndIf
    \EndFor
  \end{algorithmic}
\end{algorithm}

\subsection{Evidence Grounding}
\label{sec:grounding}

A relation passage answers the original question only through the evidence identifying its subject. In Figure~\ref{fig:problem}, the architect attribution becomes relevant through the passages identifying Bankside from the museum and year conditions. Evidence Grounding preserves these recorded connections when assembling the final generation context.

For an accepted subquestion $i\in R^t$, cumulative sources combine its direct citations with those inherited from accepted parents
\[
  C_i^t=S_i^t\cup\bigcup_{j\in\Pa(i)\cap R^t} C_j^t.
\]
Accumulation follows the DAG order. An unresolved parent contributes no inherited citations, so the child's direct citations supply the required identity connection. These source chains connect each intermediate relation to its identifying premises.

For final generation, omit $t$ and let $D\subseteq W$ be the selected passage identifiers. The eligible intermediate answers are indexed by
\begin{equation}
  I(D)=\{i\in R\mid C_i\subseteq D\}.
  \label{eq:eligible}
\end{equation}
The sequence $A(D)$ presents their tasks, answers, and citations in dependency order. Eligibility checks complete recorded citations, preserving the source context of each included intermediate result. Let $L(D)$ count the tokens in the complete generation prompt containing $Q$, instructions, passages in $D$, and $A(D)$, with input budget $B$.

When the union of cumulative sources exceeds the budget, selection considers the family $F$ of unions formed from subsets of these source sets, including the empty union. With $B\ge L(\varnothing)$, exact selection uses
\begin{equation}
  D_0\in\operatorname*{arg\,max}_{\substack{D\in F\\L(D)\le B}}
    |I(D)|.
  \label{eq:selection}
\end{equation}
The objective allocates context to intermediate answers whose recorded sources can be included together. Ties prefer smaller complete prompts, and larger Web graphs use the greedy selection described in Appendix~\ref{app:context}. Remaining capacity admits additional archived passages, with eligibility recomputed after each addition. The language model uses the selected text and $A(D)$ to connect entity identification with relation evidence and generate the answer to $Q$. Algorithm~\ref{alg:grounding} summarizes this process.

\begin{algorithm}[H]
  \caption{Evidence Grounding}
  \label{alg:grounding}
  \begin{algorithmic}[1]
    \Require Question $Q$, graph $G$, accepted answers with sources, passages $W$, budget $B$
    \Ensure Answer to $Q$
    \State Accumulate each answer's sources $C_i$ along $G$
    \State Select $D$ under the objective in Equation~\ref{eq:selection}
    \State Add passages from $W$ while preserving $L(D)\le B$
    \State Form $A(D)$ from answers with $C_i\subseteq D$ in dependency order
    \State \Return the answer generated from $Q$, $D$, and $A(D)$
  \end{algorithmic}
\end{algorithm}

\section{Experiments}
\label{sec:experiments}

We address four research questions.
\begin{itemize}
\item \textbf{RQ1.} How does \pcg{} compare with competitive baselines on complex Web QA tasks?
\item \textbf{RQ2.} How do the three components contribute to \pcg{}'s overall performance?
\item \textbf{RQ3.} How does progressive compositional grounding improve complex Web QA?
\item \textbf{RQ4.} Is \pcg{} a computationally efficient method?
\end{itemize}

\subsection{Experimental Setup}
\label{sec:setup}

\paragraph{\normalfont\bfseries Datasets.}
We evaluate complex Web QA on BrowseComp-Plus~\cite{browsecompplus} and WebWalkerQA~\cite{webwalker}. Both require identifying entities from descriptive clues and answering questions about their relations. Together, they examine this connection across a fixed Web collection and browsable pages. MuSiQue~\cite{trivedi2022musique} provides annotated supporting passages for the evidence-discovery analysis in Section~\ref{sec:mechanism}.

\paragraph{\normalfont\bfseries Metrics.}
We report judged answer accuracy on both Web QA datasets. Answer and evidence scores are percentages. Bold marks the best score and underlining the strongest baseline.

\paragraph{\normalfont\bfseries Baselines.}
The baselines are grouped into two categories. \textbf{Web search agents} include WebThinker~\cite{webthinker}, Search-o1~\cite{searcho1}, Q2Q~\cite{q2q}, and DeepHone~\cite{deephone}. \textbf{RAG} includes RADAR~\cite{radar}, LQT~\cite{lqt}, and LogicRAG~\cite{chen2026logicrag}. These groups compare complete Web-question solving with methods that explicitly organize retrieved evidence for reasoning. For evidence discovery on MuSiQue, we additionally compare with MGranRAG~\cite{hu2026mgrangrag}, CIRAG~\cite{wei2026cirag}, and ReWOO~\cite{xu2023rewoo}. Their evidence construction provides a direct comparison for recovering the premises of a compositional answer.

\paragraph{\normalfont\bfseries Implementation.}
We use Qwen3.8-27B as the base language model. Detailed experimental settings are provided in Appendix~\ref{app:implementation}.

\subsection{RQ1. Main Results}
\label{sec:mainresults}

Table~\ref{tab:main} compares \pcg{} with Web search agents and RAG methods on complex Web QA.
On BrowseComp-Plus, \pcg{} achieves 71.33\% accuracy, exceeding WebThinker by 5.43 percentage points. On WebWalkerQA, \pcg{} achieves 64.56\% accuracy, exceeding DeepHone's 59.41\% by 5.15 percentage points. The strongest baseline changes between the datasets, while \pcg{} leads in both evidence environments.

\begin{table}[!htbp]
  \centering
  \caption{Answer accuracy on complex Web QA.}
  \label{tab:main}
  \tablestyle
  \begin{tabularx}{\linewidth}{Lrr}
    \toprule
    \rowcolor{TableHead}
    \tablehead{Method} & \tablehead{BrowseComp-Plus} & \tablehead{WebWalkerQA} \\
    \midrule
    \rowcolor{TableGroup}
    \multicolumn{3}{l}{\tablehead{Web search agents}} \\
    WebThinker & \underline{65.90} & 53.68 \\
    Search-o1 & 65.30 & 55.15 \\
    Q2Q & 58.31 & 54.85 \\
    DeepHone & 55.30 & \underline{59.41} \\
    \midrule
    \rowcolor{TableGroup}
    \multicolumn{3}{l}{\tablehead{RAG}} \\
    RADAR & 31.69 & 44.12 \\
    LQT & 60.00 & 56.76 \\
    LogicRAG & 53.98 & 52.21 \\
    \midrule
    \rowcolor{TableEmphasis}
    \textbf{\pcg{}} & \textbf{71.33} & \textbf{64.56} \\
    \bottomrule
  \end{tabularx}
\end{table}

Among the RAG baselines, LQT performs best on both datasets. \pcg{} exceeds it by 11.33 and 7.80 points, respectively. LQT already organizes intermediate information needs through logical dependencies. \pcg{} strengthens this reasoning by using newly acquired identifying evidence to establish the entities filling those dependencies and connect their relations to the original question. Partial matches can thus advance retrieval before identification is complete, while subsequent sources determine which intermediate answers support dependent reasoning. The gains over both baseline groups motivate examining how acquisition, dependent reasoning, and final generation contribute to this connection.

\subsection{RQ2. Ablation Study}
\label{sec:ablation}

Table~\ref{tab:ablation} reports answer accuracy after removing each component. Appendix~\ref{app:evaluation} specifies the ablations.
\textbf{Compositional Reasoning.} Removing this component causes the largest accuracy loss on both datasets, with decreases of 13.38 percentage points on BrowseComp-Plus and 6.91 on WebWalkerQA. Question decomposition remains available, but intermediate answers no longer instantiate dependent queries, so each subquestion searches without the entities identified by its parents. The loss beyond decomposition highlights how intermediate identifications specify subsequent relation searches. \textbf{Progressive Evidence Acquisition.} Limiting acquisition to the initial retrieval lowers accuracy by 4.34 and 5.00 points. Dependent queries remain available, but initial retrieval can leave identifying conditions unresolved. Continued acquisition seeks these missing premises as intermediate answers specify new information needs. \textbf{Evidence Grounding.} Removing this component reduces accuracy by 4.70 and 5.15 points while leaving dependent queries available. A relation passage can establish a fact without establishing that its subject matches the question. Combining that passage with identifying sources and supported intermediate answers enables final generation to use the relation in the context of the original conditions.

\begin{table}[!htbp]
  \centering
  \caption{Contribution of each \pcg{} component.}
  \label{tab:ablation}
  \tablestyle
  \begin{tabularx}{\linewidth}{Lrr}
    \toprule
    \rowcolor{TableHead}
    \tablehead{Variant} & \tablehead{BrowseComp-Plus} & \tablehead{WebWalkerQA} \\
    \midrule
    \rowcolor{TableEmphasis}
    \textbf{\pcg{}} & \textbf{71.33} & \textbf{64.56} \\
    w/o Progressive Evidence Acquisition & 66.99 & 59.56 \\
    w/o Compositional Reasoning & 57.95 & 57.65 \\
    w/o Evidence Grounding & 66.63 & 59.41 \\
    \bottomrule
  \end{tabularx}
\end{table}

The larger composition loss on BrowseComp-Plus is consistent with the need to locate relation subjects from descriptive clues in a large Web corpus. Intermediate identifications make these searches specific. WebWalkerQA additionally supplies entry-page context, while relevant evidence must be found through search and navigation. Its closer ablation losses reflect the contribution of each stage, from discovering identifying passages to combining them with later relations. The two evidence stages meet at the dependent query. Acquisition supplies the premises needed to identify its subject, and grounding makes those same premises available when the retrieved relation is used in the final answer.

\FloatBarrier
\subsection{RQ3. Mechanism Analysis}
\label{sec:mechanism}

\paragraph{\normalfont\bfseries How effectively does \pcg{} discover evidence for compositional reasoning?}
Compositional reasoning requires evidence linking intermediate entities to dependent relations. MuSiQue~\cite{trivedi2022musique} pairs two to four dependent reasoning steps with annotated supporting passages for measuring complete support recovery. Table~\ref{tab:discovery} evaluates this capability against competitive RAG methods, including MGranRAG, CIRAG, RADAR, LQT, and ReWOO. Recall@5 (R@5) measures annotated support coverage, while All@5 requires the complete support set. Exact match (EM) and token F1 measure answer quality. Appendix~\ref{app:implementation} provides the settings.

\begin{table}[!htbp]
  \centering
  \caption{Evidence discovery and answer quality on MuSiQue.}
  \label{tab:discovery}
  \tablestyle
  \begin{tabularx}{\linewidth}{Lrrrr}
    \toprule
    \rowcolor{TableHead}
    \tablehead{Method} & \tablehead{R@5 $\uparrow$} & \tablehead{All@5 $\uparrow$} & \tablehead{EM $\uparrow$} & \tablehead{F1 $\uparrow$} \\
    \midrule
    NV-Embed-v2 & 69.90 & 39.30 & 38.10 & 48.17 \\
    MGranRAG & \underline{80.60} & \underline{59.90} & 49.70 & 58.63 \\
    CIRAG & 78.57 & 53.40 & 46.50 & 57.13 \\
    RADAR & 77.95 & 53.00 & \underline{52.30} & \underline{62.75} \\
    LQT & 73.98 & 46.40 & 38.00 & 48.18 \\
    ReWOO & 69.90 & 39.30 & 40.10 & 52.97 \\
    \midrule
    \rowcolor{TableEmphasis}
    \textbf{\pcg{}} & \textbf{83.91} & \textbf{64.30} & \textbf{55.70} & \textbf{65.76} \\
    \bottomrule
  \end{tabularx}
\end{table}

\pcg{} achieves 83.91\% R@5 and 64.30\% All@5, exceeding MGranRAG by 3.31 and 4.40 percentage points. An increase in R@5 can add support while leaving a reasoning chain incomplete. All@5 requires closing the final evidence gap. Their joint improvement shows broader coverage and more complete recovery of the premises connecting intermediate entities to dependent relations. This coverage is useful when answering requires several linked premises. Missing one identifying passage can disconnect a retrieved relation from the entity described by the question. MGranRAG provides the highest baseline support coverage, whereas RADAR achieves the highest baseline EM and F1. \pcg{} reaches 55.70\% EM and 65.76\% F1, exceeding RADAR by 3.40 and 3.01 points. The different baseline rankings distinguish evidence discovery from its use in answering.

Table~\ref{tab:discovery-hops} reports All@5 across hop counts, comparing \pcg{} with the four strongest baselines by overall All@5 in Table~\ref{tab:discovery}. \pcg{} achieves the highest All@5 at each depth, with gains over MGranRAG of 2.70, 7.28, and 4.22 points at two, three, and four hops. The largest gain occurs at three hops. At four hops, All@5 falls to 16.87\% despite the remaining advantage over MGranRAG's 12.65\%. The advantage in complete support recovery persists across depths, while recovering every premise becomes increasingly difficult as additional relations must be established.

\begin{table}[!htbp]
  \centering
  \caption{All@5 by reasoning depth on MuSiQue.}
  \label{tab:discovery-hops}
  \tablestyle
  \begin{tabularx}{\linewidth}{Lrrr}
    \toprule
    \rowcolor{TableHead}
    \tablehead{Method} & \tablehead{2 hops $\uparrow$} & \tablehead{3 hops $\uparrow$} & \tablehead{4 hops $\uparrow$} \\
    \midrule
    MGranRAG & \underline{81.47} & \underline{49.37} & \underline{12.65} \\
    CIRAG & 77.80 & 38.61 & 5.42 \\
    RADAR & 75.48 & 42.41 & 3.01 \\
    LQT & 71.43 & 28.80 & 1.81 \\
    \midrule
    \rowcolor{TableEmphasis}
    \textbf{\pcg{}} & \textbf{84.17} & \textbf{56.65} & \textbf{16.87} \\
    \bottomrule
  \end{tabularx}
\end{table}

\begin{figure}[H]
  \centering
  \includegraphics[width=\linewidth]{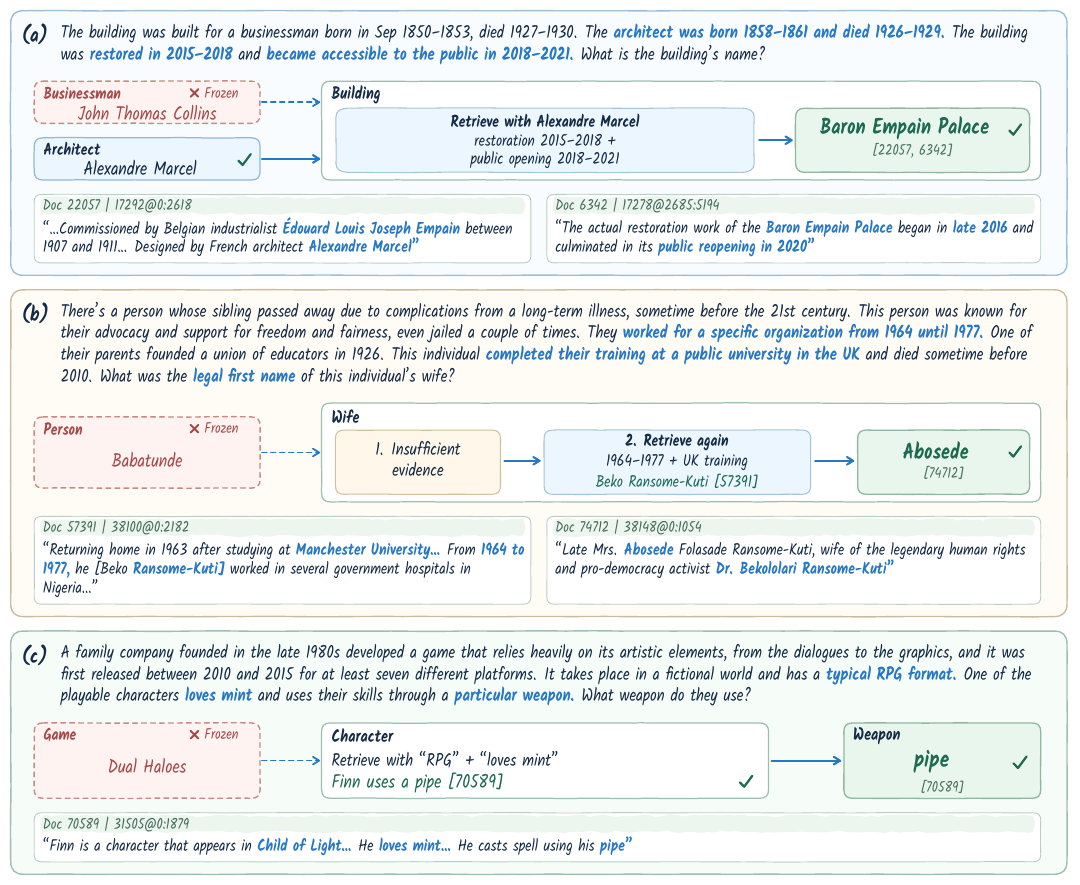}
  \caption{New evidence connects identifying clues to correct downstream answers on BrowseComp-Plus. The displayed early entity records remain unchanged.}
  \Description{Three trajectories showing an unchanged erroneous upstream entity and evidence supporting correct downstream answers. The building case combines an architect node and restoration evidence to identify Baron Empain Palace. The wife case obtains further evidence after an insufficient assessment and answers Abosede. The game case connects Finn's mint preference to his use of a pipe.}
  \label{fig:trajectory-recovery}
\end{figure}

\paragraph{\normalfont\bfseries How does new evidence support dependent reasoning after an entity error?}
Figure~\ref{fig:trajectory-recovery} shows correct downstream answers despite unchanged early entity errors. In (a), sources connect Alexandre Marcel and the restoration dates to Baron Empain Palace. In (b), the Manchester University and hospital employment clues identify Beko Ransome-Kuti, whose wife's name is given as Abosede. In (c), Finn's page connects the mint clue to the requested weapon, a pipe. The original question $Q$ accompanies each dependent task, allowing new sources to establish the intended identity and relation together. Equation~\ref{eq:bindings} also preserves unresolved identifying descriptions. Using the original clues keeps downstream answer formation connected to the intended subject. The passages establishing this connection also support final generation, grounding the relation answer in evidence about the entity described by the question.

\paragraph{\normalfont\bfseries How does evidence grounding connect entity identification to final answers?}
Table~\ref{tab:joint-support} examines final generation on 27 BrowseComp-Plus dependency questions with complete identity-to-relation support in the shared retrieved evidence. All combines the single-page and cross-page groups. Under the same context budget, relevance-based RAG selects passages independently by relevance, while \pcg{} organizes identifying and relation evidence with supported intermediate answers. Joint support measures whether the generation context establishes this connection, and accuracy measures answer correctness. Appendix~\ref{app:evaluation} details the comparison.

\begin{table}[!htbp]
  \centering
  \caption{Identifying evidence in answer generation.}
  \label{tab:joint-support}
  \tablestyle
  \begin{tabularx}{\linewidth}{lLrr}
    \toprule
    \rowcolor{TableHead}
    \tablehead{Evidence} & \tablehead{Method} & \tablehead{Joint support} & \tablehead{Accuracy} \\
    \midrule
    Single page & Relevance-based RAG & 100.0 & 83.3 \\
    & \pcg{} & 100.0 & 91.7 \\
    \midrule
    Cross page & Relevance-based RAG & 93.3 & 93.3 \\
    & \pcg{} & 100.0 & 100.0 \\
    \midrule
    \rowcolor{TableGroup}
    All & Relevance-based RAG & 96.3 & 88.9 \\
    \rowcolor{TableEmphasis}
    & \textbf{\pcg{}} & 100.0 & 96.3 \\
    \bottomrule
  \end{tabularx}
\end{table}

\begin{figure*}[!t]
  \centering
  \includegraphics[width=\textwidth]{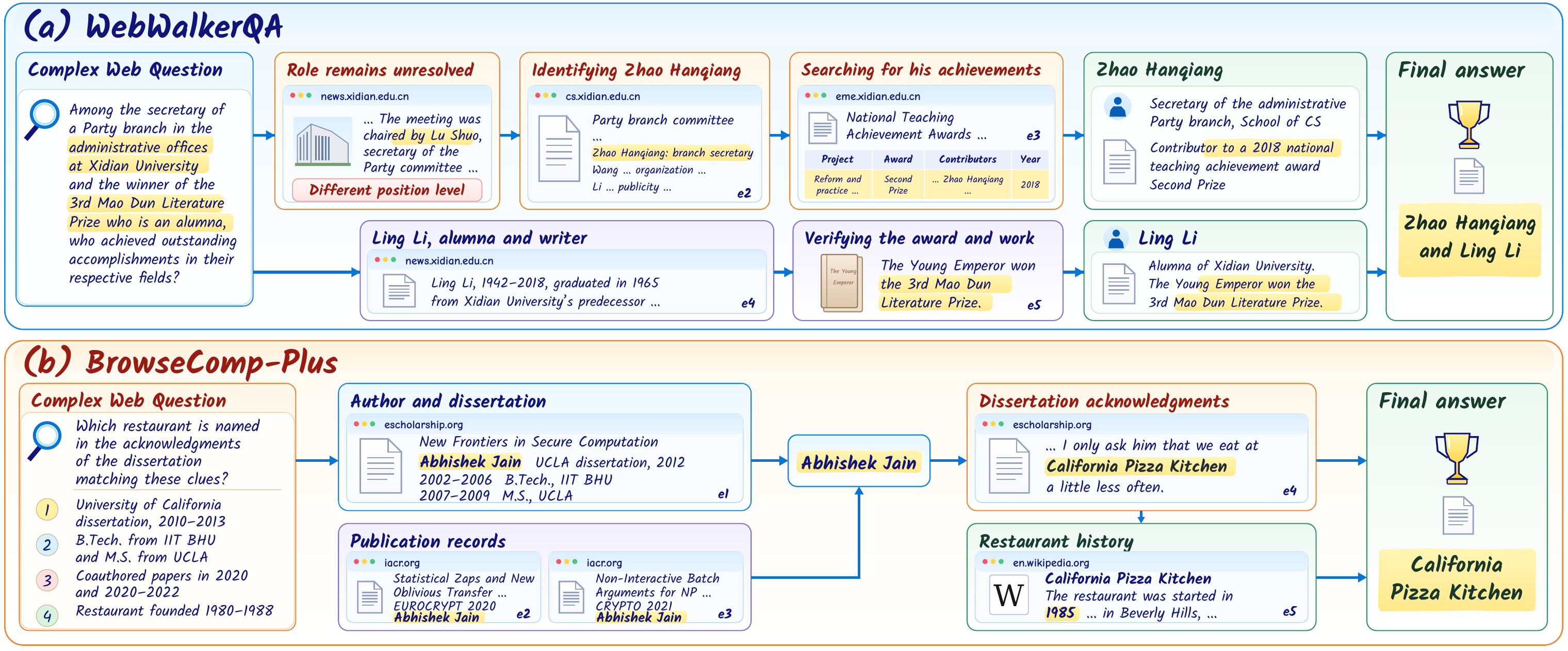}
  \caption{Identifying evidence guides subsequent relation retrieval in WebWalkerQA (a) and connects cross-page sources for final answering in BrowseComp-Plus (b).}
  \Description{Panel a distinguishes a university Party committee position from the requested administrative Party branch role, identifies Zhao Hanqiang and his teaching award, and verifies Ling Li's alumni status and literary award. Panel b combines dissertation and publication evidence to identify Abhishek Jain, then connects his dissertation's acknowledgment of California Pizza Kitchen with the restaurant's founding date.}
  \label{fig:cases}
\end{figure*}

In the cross-page group of Table~\ref{tab:joint-support}, joint support and accuracy both increase from 93.3\% to 100.0\%. Identifying passages connect the subject of a relation found on another page to the entity described by the question. Independent relevance ranking scores passages against the question individually. An identifying passage can become useful through the relation it connects, making the dependency between passages relevant to context selection. This is the connection illustrated in Figure~\ref{fig:problem}, where the museum and opening-year evidence makes the architect attribution relevant. In the single-page group, joint support is already 100.0\% in both conditions, while accuracy rises from 83.3\% to 91.7\%. Here, the difference concerns how available premises are used to produce the answer. A complete context still requires the generator to connect its sources to the conclusion. Presenting intermediate answers with the passages establishing their subjects makes that connection explicit. Across these questions, joint support increases from 96.3\% to 100.0\% and accuracy from 88.9\% to 96.3\% under the combined source-selection and answer-presentation setting. Figure~\ref{fig:cases} connects these evidence roles across complete Web QA trajectories, discussed in Section~\ref{sec:cases}.

\subsection{RQ4. Efficiency Analysis}
\label{sec:efficiency}

\paragraph{\normalfont\bfseries Inference Cost.}
Table~\ref{tab:efficiency} compares token use, model calls, and elapsed time per question. On BrowseComp-Plus, \pcg{} uses $47.40\times10^3$ total tokens and 8.00 calls, reducing token consumption by 82.98\% and latency by 62.31\% relative to WebThinker. On WebWalkerQA, its $8.11\times10^3$ tokens and 5.50 calls reduce token use by 70.86\% and latency by 51.31\% relative to DeepHone. These are the strongest accuracy baselines in Table~\ref{tab:main}, where \pcg{} also achieves higher accuracy. Appendix~\ref{app:evaluation} gives the measurement details.

\begin{table}[!htbp]
  \centering
  \caption{Computational cost of answering Web questions.}
  \label{tab:efficiency}
  \tablestyle
  \renewcommand{\arraystretch}{1.0}
  \begin{tabularx}{\linewidth}{Lrrr}
    \toprule
    \rowcolor{TableHead}
    \tablehead{Method} & \tablehead{Input / Output ($10^3$)} & \tablehead{Calls} & \tablehead{Time (s)} \\
    \midrule
    \rowcolor{TableGroup}
    \multicolumn{4}{l}{\tablehead{BrowseComp-Plus}} \\
    WebThinker & \tokenpair{265.61}{12.88} & 56.00 & 1591.23 \\
    Search-o1 & \tokenpair{138.11}{3.99} & 30.67 & 842.14 \\
    Q2Q & \tokenpair{64.37}{0.44} & 9.67 & 314.69 \\
    DeepHone & \tokenpair{55.40}{0.33} & 4.00 & 369.07 \\
    RADAR & \tokenpair{26.53}{0.79} & 13.00 & 172.63 \\
    LQT & \tokenpair{312.47}{16.92} & 64.00 & 836.52 \\
    LogicRAG & \tokenpair{43.77}{7.47} & 8.33 & 601.04 \\
    \rowcolor{TableEmphasis}
    \textbf{\pcg{}} & \tokenpair{43.10}{4.30} & 8.00 & 599.80 \\
    \midrule
    \rowcolor{TableGroup}
    \multicolumn{4}{l}{\tablehead{WebWalkerQA}} \\
    WebThinker & \tokenpair{16.44}{1.18} & 7.00 & 80.79 \\
    Search-o1 & \tokenpair{17.62}{0.77} & 5.67 & 55.45 \\
    Q2Q & \tokenpair{31.40}{0.51} & 6.00 & 35.60 \\
    DeepHone & \tokenpair{27.26}{0.57} & 4.67 & 48.16 \\
    RADAR & \tokenpair{10.09}{0.32} & 13.00 & 25.36 \\
    LQT & \tokenpair{41.94}{0.73} & 25.00 & 73.45 \\
    LogicRAG & \tokenpair{7.90}{0.26} & 3.00 & 22.12 \\
    \rowcolor{TableEmphasis}
    \textbf{\pcg{}} & \tokenpair{7.85}{0.26} & 5.50 & 23.45 \\
    \bottomrule
  \end{tabularx}
\end{table}

\paragraph{\normalfont\bfseries Cost and Effectiveness.}
LogicRAG provides a comparison at similar token budgets. \pcg{} and LogicRAG use $47.40\times10^3$ and $51.24\times10^3$ tokens on BrowseComp-Plus, with 8.00 and 8.33 calls. Their WebWalkerQA costs are $8.11\times10^3$ and $8.16\times10^3$ tokens, with 5.50 and 3.00 calls, respectively. Table~\ref{tab:main} reports accuracy advantages of 17.35 and 12.35 percentage points for \pcg{}. These results connect improved answering to how the inference process uses evidence at comparable observed token budgets.

The relation between calls and cost also varies across methods. On BrowseComp-Plus, RADAR uses fewer tokens and less time despite making 13.00 calls, compared with \pcg{}'s 8.00. Its accuracy is 31.69\%, compared with 71.33\% for \pcg{}. The higher accuracy accompanies greater token use and latency despite fewer calls. \pcg{} focuses acquisition on unresolved conditions and carries source-supported intermediate results into dependent reasoning and final generation. Input accounts for most of \pcg{}'s token use on both datasets, making the volume of evidence processed across calls an important part of inference cost. On WebWalkerQA, \pcg{} has the lowest total token use, while LogicRAG is slightly faster at 22.12 seconds compared with 23.45. Token use measures model interaction, while end-to-end latency also includes obtaining the evidence needed for those interactions.

\subsection{Case Study}
\label{sec:cases}
Figure~\ref{fig:cases} traces how identifying evidence guides dependent retrieval and supports final answers in complete WebWalkerQA and BrowseComp-Plus examples. In Figure~\ref{fig:cases}(a), the initial university report describes a Party committee position, leaving the requested administrative Party branch role unresolved. Evidence for the specific role identifies Zhao Hanqiang and directs the next search toward his teaching achievement. The parallel branch identifies Ling Li through the alumni evidence and connects her to the literary award. The final answer combines achievements tied to the people satisfying the respective identifying conditions. The example shows why a related organizational title is insufficient to determine the subject of an achievement query.

In Figure~\ref{fig:cases}(b), the university, dissertation, and publication clues establish Abhishek Jain as the intended author. His dissertation acknowledgments name California Pizza Kitchen, and the restaurant's history supplies the founding-date evidence. The acknowledgment becomes an answer to the original question through the sources identifying the dissertation. Preserving those sources with the restaurant evidence lets final generation connect the identifying constraints to the requested name across pages.

\section{Conclusion}
\label{sec:conclusion}

This paper presents \pcg{}, a novel RAG framework that links entity identification with compositional reasoning in complex Web QA. By acquiring missing identifying evidence through Progressive Evidence Acquisition and instantiating dependent queries through Compositional Reasoning, our method uses evolving entity judgments to guide retrieval while preserving the original question's identifying conditions. The Evidence Grounding module further aligns final generation with the question by combining identifying and relation sources with the intermediate answers they jointly support. The superior results of \pcg{} across three QA benchmarks validate that progressive compositional grounding can improve retrieval-augmented answering by connecting entity identification with evidence-supported answer generation.
\label{sec:main-end}

\bibliographystyle{ACM-Reference-Format}
\bibliography{references}
\appendix
\raggedbottom
\section{Dataset and Retrieval Settings}
\label{app:settings}

Table~\ref{tab:datasets} summarizes the evaluation data. All methods receive the same questions and evidence resources within each dataset.

\begin{table}[H]
  \centering
  \caption{Evaluation datasets and evidence environments.}
  \label{tab:datasets}
  \tablestyle
  \begin{tabularx}{\linewidth}{lrL}
    \toprule
    \rowcolor{TableHead}
    \tablehead{Dataset} & \tablehead{Questions} & \tablehead{Evidence environment} \\
    \midrule
    BrowseComp-Plus & 830 & Shared Web corpus \\
    WebWalkerQA & 680 & Search and navigation \\
    MuSiQue & 1,000 & Wikipedia passages \\
    \bottomrule
  \end{tabularx}
\end{table}

\appparagraph{BrowseComp-Plus.}
We evaluate 830 questions on a shared corpus snapshot with stable page identifiers. Each query combines the top 200 dense candidates with the top 200 BM25 candidates through reciprocal rank fusion ($k=60$). After deduplication, the shared reranker processes at most 400 pages. Dense retrieval uses Qwen3-Embedding-0.6B. Qwen3-Reranker-0.6B ranks query-relevant windows of up to 768 tokens~\cite{qwenembedding}.

\appparagraph{WebWalkerQA.}
We use the official 680-question main set and its entry URLs. Each Exa query returns up to ten unique accessible results, and open search can discover pages beyond the entry site. Navigation follows observed links under shared filtering and page-access policies.

\appparagraph{MuSiQue.}
The evidence-discovery analysis uses 1,000 questions and 11,656 Wikipedia passages. Annotated supporting passages are used only for evaluation.

\section{Implementation Details}
\label{app:implementation}

\subsection{Baseline Configurations}
\label{app:baselines}

WebThinker uses its native Explorer controller, and Search-o1 keeps its search and Reason-in-Documents workflow. Both access the shared retrieval and page-access services. RADAR, LQT, and Q2Q use their method-specific reasoning procedures. DeepHone uses the shared Qwen3-Reranker-0.6B cross-encoder. LogicRAG uses its native adaptive controller and generation procedure. Its BrowseComp-Plus configuration permits three additional retrieval rounds with ten pages per round.

For MuSiQue, all methods use Qwen3.8-27B and NV-Embed-v2 embeddings~\cite{lee2025nvembed}. We compute R@5 and All@5 from each method's native evidence ranking. MGranRAG uses its final passage ranking. CIRAG uses fact-derived rankings with its standard fallback. Original-question retrieval provides the dense reference in Table~\ref{tab:discovery}.

\subsection{PCG-RAG Configuration}
\label{app:context}

The experiments run on a server with eight NVIDIA GeForce RTX 3090 GPUs. The Web experiments use Qwen3.8-27B with AWQ-INT4 weights, greedy decoding, and hidden thinking disabled. Table~\ref{tab:settings} reports input and base output budgets in tokens. Follow-up actions are counted per subquestion. Input budgets cover the complete prompt, including the task context and tool interface.

\begin{table}[H]
  \centering
  \caption{PCG-RAG inference budgets.}
  \label{tab:settings}
  \tablestyle
  \begin{tabularx}{\linewidth}{Lrr}
    \toprule
    \rowcolor{TableHead}
    \tablehead{Setting} & \tablehead{BrowseComp-Plus} & \tablehead{WebWalkerQA} \\
    \midrule
    Planner output & 2,048 & 4,096 \\
    Subquestion input & 8,192 & 8,192 \\
    Subquestion output & 512 & 1,024 \\
    Generation input & 12,288 & 12,288 \\
    Generation output & 1,024 & 4,096 \\
    Follow-up actions & 4 & 8 \\
    \bottomrule
  \end{tabularx}
\end{table}

\appparagraph{Acquisition bounds.}
Each attempt starts with retrieval. Admissible follow-up actions acquire new evidence using unused queries, observed links to accessible pages, and unseen reading windows. Acceptance ends the attempt. Further acquisition requires both remaining allowance and an admissible action.

\appparagraph{Context selection.}
When the union of cumulative sources exceeds the budget, plans with at most ten accepted subquestions enumerate source-set unions under Equation~\ref{eq:selection}. Ties prefer fewer prompt tokens, then earlier passage-preference order. Larger Web plans start with $D=\varnothing$ and add feasible source sets maximizing
\[
  \frac{|I(D\cup C_i)|-|I(D)|}
       {\max\{1,L(D\cup C_i)-L(D)\}}.
\]
Greedy ties prefer greater coverage gain, then shorter input. Remaining capacity admits candidate citations, interleaved retrieval results, then archived passages, recomputing eligibility and complete prompt length after each addition.

For MuSiQue, \pcg{} uses one solving attempt per subquestion with follow-up acquisition disabled. Tables~\ref{tab:discovery} and~\ref{tab:discovery-hops} evaluate evidence discovery under this single-pass configuration. Each search retrieves 50 dense candidates, and the subquestion solver receives up to 20 passages. Final generation uses five passages for \pcg{}, MGranRAG, and the dense reference. Evidence rankings use a common cutoff for R@5 and All@5. CIRAG uses its adaptive generation procedure. ReWOO generates answers from its plan observations.

\section{LLM Prompts}
\label{app:prompts}

The instructions connect evidence acquisition to the original question's identifying conditions. Dependent answers require citations that establish the described entity.

\begin{promptcard}{Question Decomposition}
\promptfield{Input}
The original question \(Q\).

\promptfield{Prompt}
Plan the evidence needed to answer the original question completely. Preserve the full answer specification across the plan. List steps in dependency order. Each step names only the earlier output slots that it needs. Reference their answers in question as \promptvar{\{slot\}}. Use concise executable questions. Output only the JSON object.

\promptfield{Output}
A JSON plan with ordered subquestions and explicit dependencies.
\end{promptcard}

\begin{promptcard}{Progressive Evidence Acquisition}
\promptfield{Input}
The original question with the current task and acquisition state.

\promptfield{Prompt}
Use \promptvar{status=supported} when the selected sources establish the complete answer to the current task. Use \promptvar{status=provisional} when a plausible candidate has unresolved identifying constraints. Record the candidate in \promptvar{answer}. Cite its identifying passages in \promptvar{sources}. Record the unresolved constraint in \promptvar{missing}. Direct \promptvar{next\_queries} toward this constraint.

\promptfield{Output}
A status judgment with cited evidence and follow-up queries.
\end{promptcard}

\begin{promptcard}{Compositional Reasoning}
\promptfield{Input}
A dependent task with parent results and source passages.

\promptfield{Prompt}
Parent hypotheses are search seeds. For a dependent answer, the sources must establish the requested fact about the intended entity. They must also connect that entity to the original question. If the identity remains uncertain, keep the dependent answer provisional.

\promptfield{Output}
A dependent answer with its support status and cited evidence.
\end{promptcard}

\begin{promptcard}{Evidence Grounding}
\promptfield{Input}
The original question with selected sources and eligible intermediate answers.

\promptfield{Prompt}
Answer the original question using the supplied source passages. Include every supported answer component. Preserve the requested answer scope and all source constraints. Treat intermediate answers as working proposals. Return a JSON object containing exactly one string field named \promptvar{answer}.

\promptfield{Output}
A JSON object with one string field named \promptvar{answer}.
\end{promptcard}

\section{Evaluation Protocol}
\label{app:evaluation}

\appparagraph{Answer and evidence metrics.}
DeepSeek-V4.1-Flash judges Web QA answers against each dataset's references and judging criteria. For $N$ MuSiQue questions, let $G_i$ be the annotated supporting passages~\cite{trivedi2022musique} and $D_i^5$ the set of the first five passages in the method's evidence ranking. R@5 measures passage coverage, following retrieval evaluation in MGranRAG~\cite{hu2026mgrangrag}. All@5 measures recovery of the complete support set. Let $\hat a_i$ be the predicted answer and $A_i$ the reference aliases for question $i$. EM and token F1 follow standard QA evaluation~\cite{rajpurkar2016squad}. Normalization $z$ lowercases text, removes punctuation and articles, and standardizes whitespace. Let $T(a)$ be the normalized token multiset and $o_i(a)=|T(\hat a_i)\cap T(a)|$ count overlapping tokens with multiplicity. The percentage scores are
\[
\begin{aligned}
\mathrm{R@5}&=\frac{100}{N}\sum_{i=1}^{N}\frac{|D_i^5\cap G_i|}{|G_i|},\\
\mathrm{All@5}&=\frac{100}{N}\sum_{i=1}^{N}\mathbf{1}[G_i\subseteq D_i^5],\\
\mathrm{EM}&=\frac{100}{N}\sum_{i=1}^{N}\max_{a\in A_i}\mathbf{1}[z(\hat a_i)=z(a)],\\
\mathrm{F1}&=\frac{100}{N}\sum_{i=1}^{N}\max_{a\in A_i}
\frac{2o_i(a)}{|T(\hat a_i)|+|T(a)|}.
\end{aligned}
\]
Here $\mathbf{1}$ is the indicator function. The F1 fraction is defined as zero when $o_i(a)=0$. EM and F1 take the best reference-alias score before averaging across questions.

The two-, three-, and four-hop groups contain 518, 316, and 166 questions. Table~\ref{tab:discovery-hops} applies All@5 to each group.

\appparagraph{Module ablations.}
Table~\ref{tab:ablation} keeps the backbone, datasets, retrieval resources, and generation budget fixed. Without \emph{Progressive Evidence Acquisition}, each subquestion receives initial retrieval and one assessment. Follow-up acquisition is disabled, so unresolved identifying conditions do not trigger additional searches within that attempt.

Without \emph{Compositional Reasoning}, subquestions use their evidence independently. Parent-answer substitution is disabled, removing the use of an intermediate answer to instantiate later relation queries. Without \emph{Evidence Grounding}, generation uses independently relevance-ranked passages and committed intermediate answers. Inherited-source priority and complete-source eligibility are disabled, so presenting an intermediate answer does not require its complete recorded source chain in the generation context.

\appparagraph{Identifying evidence and joint support.}
Table~\ref{tab:joint-support} compares two generation conditions on the same 27 BrowseComp-Plus dependency questions with complete source support, using shared retrieved evidence and input budgets. Relevance-based RAG selects passages independently by relevance. \pcg{} combines identifying and relation sources with eligible intermediate answers. The comparison jointly varies source selection and intermediate-answer presentation.

Joint support requires original passages in the generation context to connect the entity described by the question with the requested relation. Single-page cases have complete identifying and relation support on one page, while cross-page cases require evidence from multiple pages. These groups contain 12 and 15 questions and together form All. Of the original 40 reviewed questions, 13 have incomplete support and fall outside this comparison. Accuracy measures final-answer correctness on the same 27 questions, with $24/27$ correct for relevance-based RAG and $26/27$ for \pcg{}.

\appparagraph{Cost measurement.}
Each method uses the same 30 questions within each dataset. Table~\ref{tab:efficiency} reports per-question arithmetic means for input tokens, output tokens, calls, and elapsed time. Counts include every actual LLM request for planning, all subquestion solving, final generation, and output retries, excluding scoring. Tokens use units of $10^3$ and time uses seconds.

\end{document}